\documentstyle[11pt,newpasp,twoside,epsf,psfig]{article}
\def\edcomment#1{\iffalse\marginpar{\raggedright\sl#1\/}\else\relax\fi}
\reversemarginpar

\begin{document}
\title{Imaging in Hard X-ray Astronomy}
 \author{T.P. Li}
\affil{Physics Dept. and Center for Astrophys., Tsinghua University, Beijing
\\Institute of High Energy Physics, Chinese Academy of Sciences, Beijing}

\begin{abstract} 
The energy range of hard X-rays is a key waveband to the study of high energy
processes in celestial objects, but still remains poorly explored. 
 In contrast to direct imaging methods used in the low energy X-ray and 
high energy gamma-ray bands, currently imaging in the hard X-ray band 
is mainly achieved through various modulation techniques. 
A new inversion technique, the direct demodulation method,
 has been developed since early 90s. 
With this technique, wide field and high resolution images can be derived 
from scanning data of a simple collimated detector. The feasibility
of this technique has been confirmed by experiment, balloon-borne observation
and analyzing simulated and real astronomical data. Based the development of 
methodology and instrumentation, a high energy astrophysics 
mission -- Hard X-ray Modulation Telescope (HXMT) has been proposed
and selected in China for a four-year Phase-A study.
The main scientific objectives are a full-sky hard X-ray (20-200 keV)
imaging survey 
and high signal-to-noise ratio timing studies of high energy sources.
\end{abstract}

\section{Introduction}
  The hard X-ray band  with photon energies from $\sim 10$~keV
to a few hundred keV is very important for the study of high energy processes 
near compact objects, super-massive black holes and in relativistic outflows. 
Astronomical imaging in hard X-rays is still 
an observational problem. This is because high energy photons in this range 
can neither be focused like lower energy photons, nor can their arrival
direction be determined from Compton scattering or e$^{\pm}$ pair
production as in the high energy $\gamma$-ray range.
 Due to the technological difficulties no detailed all-sky map of 
hard X-rays has been provided yet.  

In contrast to direct imaging methods used in the optical and 
soft X-ray bands, 
currently imaging in the hard X-ray band is mainly 
achieved through various modulation techniques.
M. Oda (1965) proposed that a collimator composed of two parallel
grids of wires, scanning cross a X-ray source over a detector, could
reveal the positional information of the source. After then 
the rotation modulation collimator (RMC) technique (Schnopper et al. 1968;
 Willmore 1970) and coded aperture mask (CAM) technique (Dick 1968;
Ables 1968) were developed with good angular 
resolution and a wide field of view (FOV) for hard X-ray imaging. 
By rotating the modulation grids, a periodic modulation
of the source flux is introduced which contains unique information
on the position and intensity of the sources in the FOV. Images
are obtained with the aid of Fourier analysis or cross-correlation
analysis. Composite-aperture telescopes, e.g. multipitch
telescopes, may be used in order to reduce the side lobes and
perform imaging of diffuse sources. Currently, in hard X-
ray imaging observations, the CAM is widely
used. Modulated spatially by the coded aperture mask, the
incident photons are
recorded by a position sensitive detector. The true spatial
intensity distribution is reconstructed from the observational
data by using the cross-correlation analysis method or other
mathematical decording techniques. 
The coded aperture mask technique is most widely used in hard X-ray imaging:  
modulated spatially by the coded aperture
mask, the incident photons are recorded by a position sensitive detector.
The true objective distribution is reconstructed from the observational data
by using the cross-correlation (CC) technique or other inversion methods.
Due to the position sensitive detector a hard X-ray coded aperture telescope  
is fairly complicated and expensive. The intrinsic angular resolution 
of a coded aperture mask instrument can be estimated by
$\Delta = \arctan (2\delta x/d)$,         
where $\delta x$ is the detector-mask pixel size and $d$ the distance between
the mask and detector plane. With a limited position resolution of the 
detector, say 1 cm, the telescope can reach a length of $6 \sim 7$ m to get 
about $10'$ angular resolution, which leads to large structure, heavy weight and 
difficulties in suppressing the local background. The effective area is reduced
to half or more due to the occultation of the mask and a wide field of view
will lead to a higher background through the aperture. 
Complicated image distortions (e.g. side lobes, 
pseudo images etc.) are usually appeared in images obtained by coded aperture
 telescopes, which strongly reduce the angular resolution and imaging capability 
for weak sources.

 Since early 1990s a direct demodulation method has been developed.
The direct demodulation technique can extract the information on the observed object
from the observational data more sufficiently than conventional inversion techniques,
 with which one can perform wide field and high resolution imaging with a 
non-position-sensitive detector and relatively simple modulation methods. 
 
\section{Direct Demodulation Method}

 In observations with any kind of telescopes, the relation between 
the observational data $d(k)$ and the intensity distribution $f(i)$ of
a sky region can be described by the following observation equation 
(modulation equation):
\begin{equation}
\sum\limits_{i=1}^{N}p(k,i)f(i)=d(k)\hspace{10mm} (k=1,...,M)   
\end{equation}
where $p(k,i)$, the modulation coefficients of the telescope, represent
the instrument response character. 
The equation system (1) can be written in matrix form as 
\begin{equation} 
       Pf = d           
\end{equation}

Since early 1990s the direct demodulation (DD) technique has been developed
(Li \& Wu 1992; 1993; 1994) for deriving high spatial resolution
maps from incomplete and noisy data. The DD technique reconstructs the object $f$ 
from the observed data $d$ by directly solving the observation equations.
In general the modulation matrix $P$ is not a square one and the observation equation 
system is unsolvable. We can first multiply two sides of Eg. (2) by the transpose, $P^{T}$,
of the modulation matrix to obtain a new equation system
\begin{equation}
 P'f = c                
\end{equation}   
where $c=P^{T}d, P'=P^{T}P$. The matrix of coefficients, $P'$, in the correlation equations (3)
is then a positive definite symmetric matrix.  
The DD technique performs a  deconvolution 
from $c$ by iteratively solving Eg. (3) under some proper physical
constraints ( Li \& Wu 1994). The formula of DD
algorithm by using the Gauss-Seidel iterations is 
\begin{equation}
 f^{(l)}(i) = \frac{1}{p'(i,i)} [c(i) - \sum_{j\neq i} p'(i,j)
f^{(l-1)}(j)]
\end{equation}
with the constraint condition
\begin{equation}
 f(i) \geq b(i) 
\end{equation}
where the lower intensity limit $b(i)$ is the background intensity. 

 The iterative process described in equation (4) under the nonlinear 
constraint (5) is in fact 
a sort of artificial neural network computing (Li 1997). Let
$x(i)=f(i)-b(i), ~\theta(i)=b(i)[1-\sum_{j\neq i}p'(i,j)/p'(i,i)]
-c(i)/p'(i,i), ~w(i,j)=-p'(i,j)/p'(i,i)$, 
the DD algorithm (4), (5) can be represented by the 
state transfer equation of a continual Hopfield network (Hopfield 1984)
\begin{equation}
 x^{(l)}(i) = F[\sum_{j\neq i} w(i,j)x^{(l-1)}(j)-\theta(i)] 
\end{equation}
where $x(i)$ is the output of neuron $i$, $\theta(i)$  the
threshold of neuron $i$, $w(i,j)$ the weight from the output of neuron $j$
to the input of neuron $i$, and
the transfer function $F$ is a nonlinear ramp function
\begin{equation}
 F(x) =  \left\{ \begin{array}{ll}
                   x   & \mbox{if $x\geq 0$} \\
                   0   & \mbox{otherwise}
                 \end{array}
         \right. 
\end{equation}     
Figure 1 shows a three-neuron network representing the DD
algorithm.  Cohen and Grossberg (1983) have shown that recurrent networks are stable
if the weight matrix is symmetrical with zeros on its main diagonal
and the inverse function of transfer function is continual and
monotonously increasing. It is easy to see that the network defined by
(6) and (7) satisfies the above mentioned conditions. Therefore the direct 
demodulation
calculations always converge to a stable solution, a globally optimal   
one in the meaning of the Liapunov energy function being minimum.    
\begin{figure}
\psfig{figure=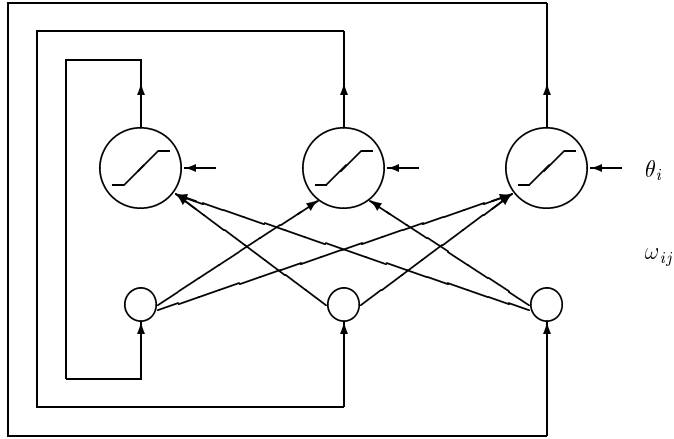,width=13cm,angle=0,%
bbllx=20pt,bblly=480pt,bburx=500pt,bbury=724pt,clip=}
\caption{Continual Hopfield Network with Three Artificial Neurons.}
\label{picture}
\end{figure}

An alternative algorithm of direct demodulation is to solve the observation equation
system (1) by Richardson-Lucy iteration (Richardson 1972; Lucy 1974)
under the background constraint:
\[
f^{(l)}(i)=f^{(l-1)}(i)\sum_j\frac{p(j,i)d(j)}{\sum_{i'}p(j,i')f^{(l-1)}(i')}/\sum_jp(j,i)
\]
\begin{equation}
f(i)\geq b(i)
\end{equation}

The DD method is a general inversion method, which can be used to deal with
observational data obtained by different kinds of instrument and has been used to 
analyze data from 
both simulations and real space experiments with  various types of telescope.
The results obtained show that the
DD technique can improve the spatial resolution and sensitivity significantly.  
Mont Carlo simulations show that, comparing with the conventional 
cross-correlation deconvolution technique the DD technique can improve
the image quality for rotation modulation telescope (Chen et al. 1998)
and coded mask telescope (Li 1995). 
Figure 2 shows the simulation results of the DD image reconstruction for a 
rotating modulation telescope (Chen et al. 1998).         
 The image reconstructed from a object shown in Fig. 2(a) with the cross-correlation 
and with the DD technique are shown in figures 2(b) and 2(c), respectively.
From figure 2(c) one can see that with the DD technique much better
reconstruction for point source, extended source and flat background 
can be obtained. 
\begin{figure}
\psfig{figure=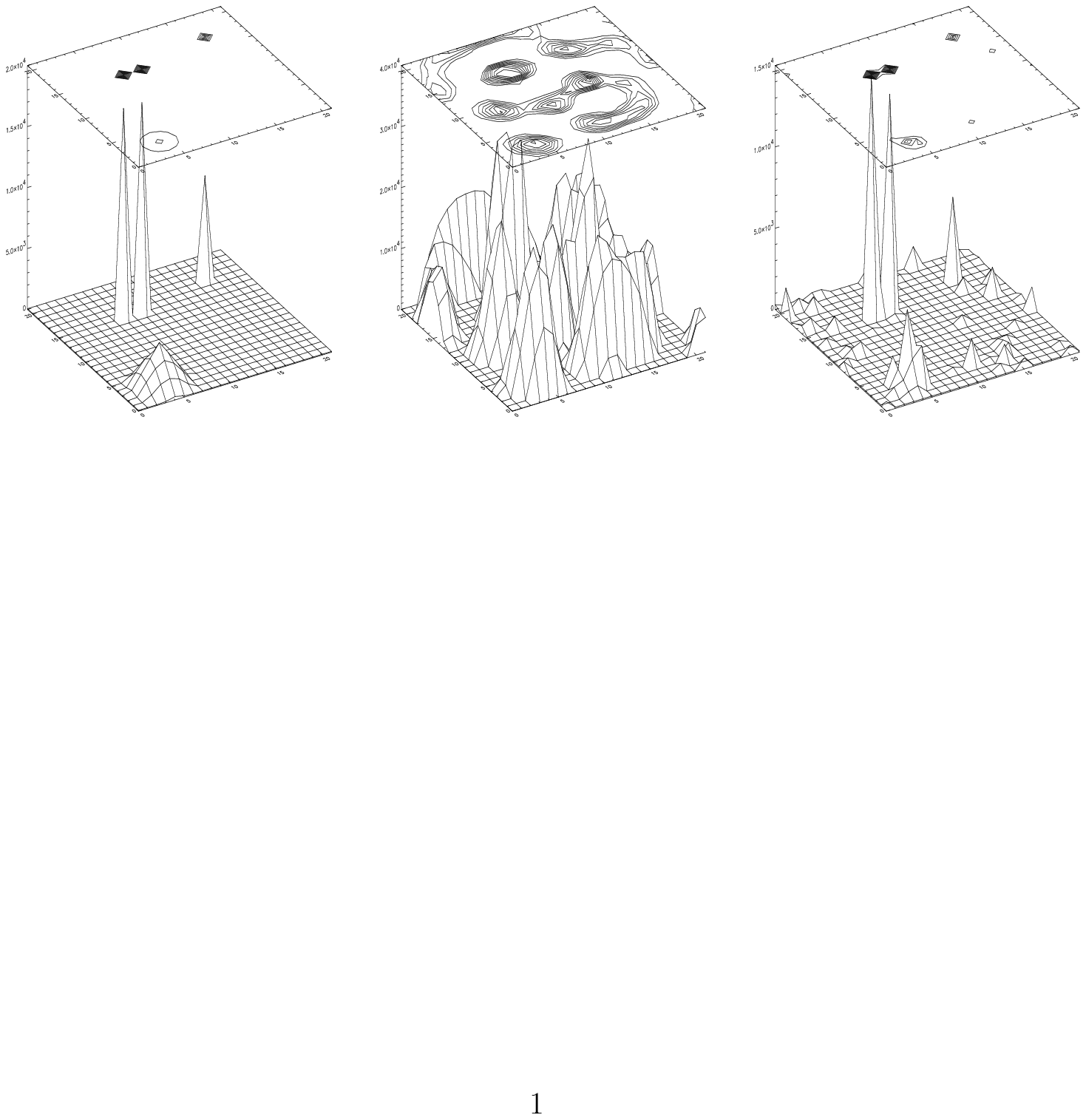,width=17truecm,angle=0,%
bbllx=90pt,bblly=380pt,bburx=600pt,bbury=564pt,clip=}
\caption{Images with a rotating modulation telescope. The coordinates are in arbitrary units.
(a)Object scene; (b) Cross-correlation; (c) Direct demodulation.}
\end{figure}

The DD technique has been successfully applied to improving the angular resolution of
the Wolter I type telescope PSPC aboard ROSAT mission (Chen et al. 1998). 
A strong evidence for an X-ray jet associated with SNR G54.1+0.3 has been found 
with PSPC/ROSAT and DD technique (Lu et al. 2001).

To construct intensity map from the Compton telescope with unusual modulation
function and 3-D data space is a difficult task in space high energy
astronomy. The DD technique has also been successfully 
applied to high resolution imaging of the double Compton scattering 
telescope COMPTEL on board CGRO. By using this technique 
Zhang et al. (1998)   re-analyzed the 
data from the COMPTEL observation of PKS0528+134 during the 1993 March 
flare in $\gamma$-rays; the result revealed new information about the flare activity 
of the Blazar. As an example of DD images from Compton telescope, figure 3 shows
the first 10-30 MeV map of the Geminga pulsar in the phase region of peak 1 
 determined by the light curve of the Geminga pulsar 
from the EGRET data in the same observing period (Zhang et al. 1997).
\begin{figure}
\psfig{figure=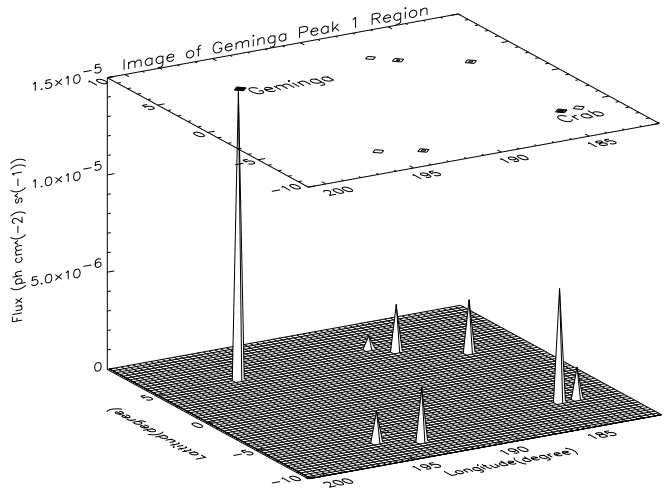,width=10cm,angle=0,%
bbllx=20pt,bblly=360pt,bburx=540pt,bbury=740pt,clip=}
\caption{The phase resolved 10-30 MeV  map of Geminga from the COMPTEL data with the DD technique.}
\label{picture}
\end{figure}
The COMPTEL observation of the $\gamma$-ray burst GRB 910601 has been re-analyzed using the DD
method (Zhang 1998). The location of GRB 910601 derived from the DD imaging is closer 
to the annulus obtained by the Ulysses-BATSE system than that from the maximum-likelihood method.
 
The all-sky monitor (ASM), aboard the Rossi X-ray Timing Explorer (RXTE),
consists of one-dimensional position-sensitive counters mounted on a motorized rotation 
drive viewing the sky through slit masks. Intensities of sources listed 
in a master catalog of X-ray sources with known positions are obtained 
via least squares fits of shadow patterns to the data (Levine et al. 1996). 
Song et al. (1999) has successfully performed image reconstruction from 
ASM/RXTE data with the DD technique without using any  a prior
knowledge about the sources. Figure 4 shows  a DD image with
ASM/RXTE data and DD technique.
\begin{figure}
\psfig{figure=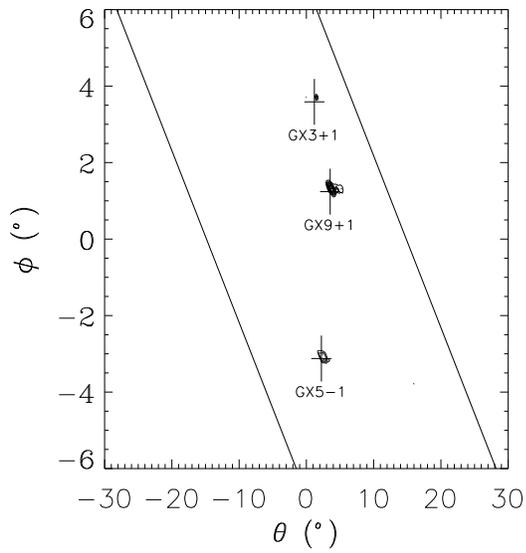,width=10cm,angle=0,%
bbllx=20pt,bblly=360pt,bburx=540pt,bbury=840pt,clip=}
\caption{ASM/RXTE images obtained with the DD technique.}
\label{picture}
\end{figure}

The DD technique has been applied to derive high resolution maps from
scanning observations by non-position-sensitive detectors.
The European X-ray satellite EXOSAT, in a survey mode, made scanning 
observation of the galactic plane with its medium energy (2-6 keV) 
collimated 
detector, ME, which had a spatial resolution of $\sim 0^{\circ}.75$ (FWHM). The 
data analysis technique used by the EXOSAT team at that time was unable to 
resolve 
some of the complex regions (Warwick et al. 1988). Lu F.J. et al. (1996)
re-analyzed the EXOSAT-ME scan data using the DD
technique and got new X-ray maps of 
the Galactic plane which are much better than those made by the traditional 
cross-correlation method. For example, the region 
around $l=346^{\circ}$ is 
a confused region in the previous analysis, but with the DD technique 
 a well resolved X-ray map is derived successfully. Figure 5(b) is the X-ray
map of this region derived with the DD technique, revealing quite a few point 
sources in an 
$8^{\circ}\times8^{\circ}$ region. In comparison, figure 5(a) is the 
cross-correlation map. All point sources in Fig. 5 observed by the DD technique
have been confirmed by X-ray imaging telescopes, including the weak source
DD 1701-380 confirmed later by 2E 1701.1-3804 found in accumulative data 
of Einstein Observatory ( Thompson et al. 1998). 
 \begin{figure}
\psfig{figure=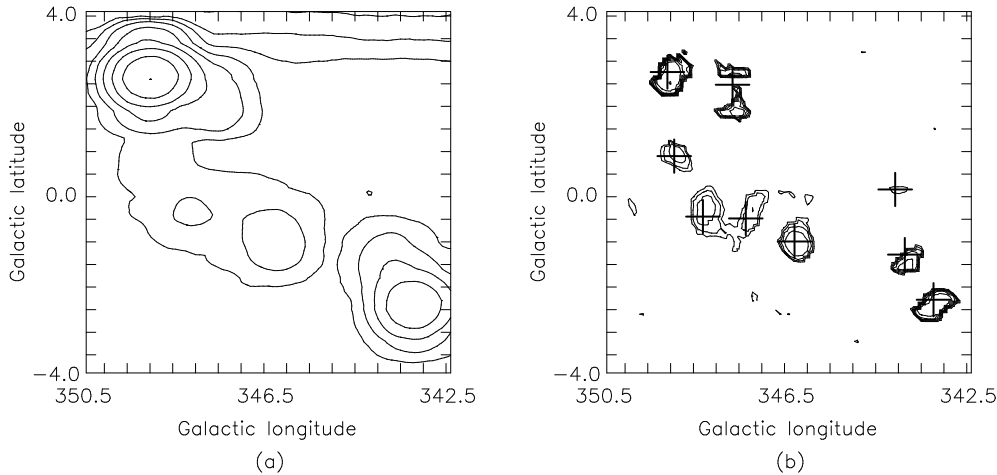,width=14cm,angle=270,%
bbllx=50pt,bblly=0pt,bburx=320pt,bbury=570pt,clip=}
\caption{X-ray map of the region around $l$=346$^{\circ}$ from EXOSAT-ME slew
observations. (a): Cross-correlation map. (b): Map derived with the DD technique.}
\label{picture}
\end{figure}
Another example is for the hard X-ray survey telescope HEAO1-A4, which consists of two similar 
detectors with 100 cm$^{2}$ sensitive area, in the 13--180 keV energy range, 
and FOV of $1^{\circ}.5\times20^{\circ}$ (FWHM) each (Matteson 1978). 
The two detectors rotate around the pointing axis clockwisely and 
counter-clockwisely by 30$^{\circ}$, respectively. The detectors scan the 
sky by the orbit precession of the satellite (Levine et al. 1984). 
No traditional technique can produce meaningful images from the HEAO1-A4 
data. However, good quality images have been 
derived with the DD technique. Figure 6(b), as an 
example, is a direct reconstruction from the scanning 
observation data of HEAO1-A4 for the Galactic center region, 
where the crosses mark the positions of the known hard X-ray sources (Lu F.J. et al. 1995). 
Again, in comparison, figure 6(a) gives the cross-correlation map of the same data.
\begin{figure}
\psfig{figure=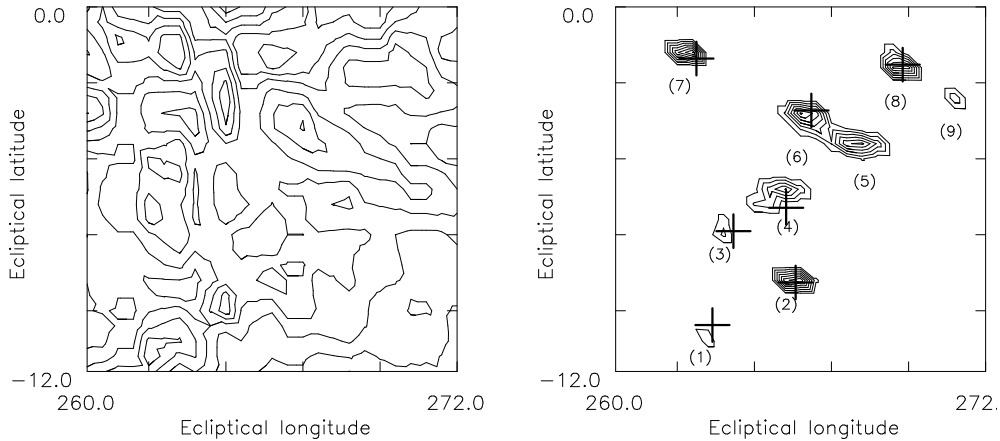,width=14cm,angle=270,%
bbllx=90pt,bblly=0pt,bburx=320pt,bbury=504pt,clip=}
\caption{Hard X-ray (13-180 keV) map of the galactic center region derived 
from HEAO1-A4 all-sky survey data. {\it Left}: Cross-correlation map. 
{\it Right}: Map derived with the DD technique.}
\label{picture}
\end{figure}

\section{The Hard X-ray Modulation Telescope HXMT}
The possibility of a collimated telescope to perform high resolution imaging
with scan observation and DD technique
(Li et al. 1993; Li \& Wu 1994) has been proved by the results of reanalyzing
 space data, shown in Figs. (5) and (6), and by raster scan imaging experiments
in the laboratory and  balloon flight 
 with the hard X-ray telescope HAPI-4 (Lu Z.G. et al. 1995). 
The balloon-borne telescope HAPI-4 is a slat-collimated hard X-ray telescope 
of the Institute of High Energy Physics 
(IHEP),Beijing, in 
cooperation with the Moscow Engineering Physics Institute of Russia and 
Astronomisches Institut, Universitaet Tuebingen of Germany.  
The telescope consists of
a collimated multiple wire proportional counter (MWPC) and NaI(Tl)/CsI(Na) phoswich 
detector. The effective area of both the MWPC and the phoswich detector is 1600 cm$^2$
each and the energy range is 10-400 keV (10-100 keV when only the MWPC are used). The
collimator is made of lead slats, which define a FOV of 3$^\circ$$\times$3$^\circ$ (FWHM).
A raster scan modulation imaging experiment for radioactive sources
in the laboratory with HAPI-4 has been done and the observed data were analyzed with the  DD
technique. The derived positions and flux of the radioactive sources are
consistent with the true values.    
On September 25, 1993, at the altitude of about 35 km over Beijing at a northern 
latitude of 40$^{\circ}$ a raster scan observation of the Cygnus region was performed
for about one hour by HAPI-4.  The
observed data were analyzed with DD technique (Lu Z.G. et al. 1995). 
The derived X-ray image of the Cygnus region is 
shown in figure 7.
\begin{figure}
\psfig{figure=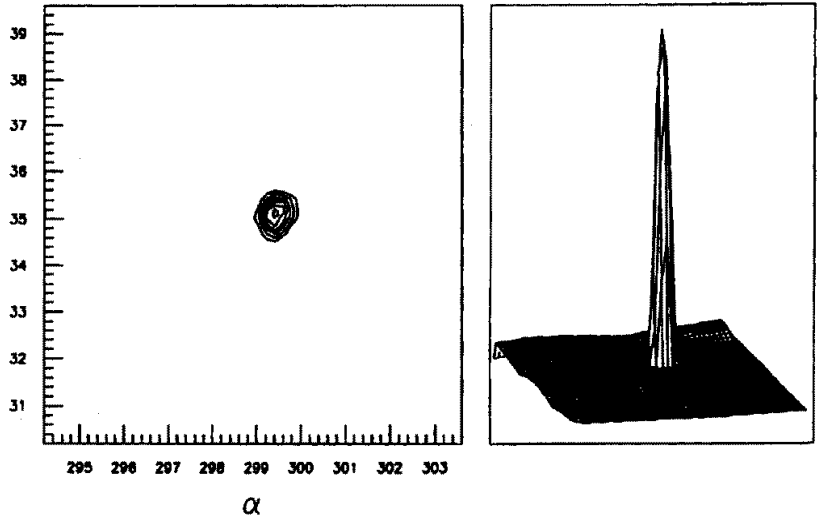,width=11cm,angle=0,%
bbllx=-180pt,bblly=50pt,bburx=850pt,bbury=840pt,clip=}
\caption{The X-ray map of the Cygnus region with Cyg X-1 in the center. It was
obtained by HAPI-4, a balloon borne hard X-ray telescope.}
\label{picture}
\end{figure}

 Based the development of 
methodology and instrumentation, a Chinese high enerrgy astrophysics 
mossion -- Hard X-ray Modulation Telescope (HXMT) has been proposed. 
In the energy range of 10 - 200 keV, HXMT can make full-sky hard X-ray 
survey with high spatial resolution, deep imaging observations of
selected sky regions, and high sensitivity pointing observations of 
scientific hot spot sources for detailed temporal and spectral studies. 

  The detector of the HXMT consists of 18 same hexagonal prism NaI(Tl)/CsI(Na) 
phoswiches (Figure  8), the area of single module is 286 cm$^{2}$, 
then the total detecting area is 5148 cm$^{2}$. 
The primary detector of each module is a NaI(Tl) crystal with the thickness of 
3 mm, and a 500 $\mu$m thick Be slice is used as the incident window.
A CsI(Na) crystal of 3 cm thick is placed at the back of the primary crystal
to act as an anticoincidence shield: it stops the hard X-ray and $\gamma$-ray 
background of the lower 2$\pi$ solid angle (assuming the telescope is pointing
upwards) and reduces the effect of Compton scattering in the primary counter.
The two kinds of crystals are optically coupled. Fluorescence photons are
received by a $5''$ photomultiplier (PMT) for each phoswich module.
A pulse shape discrimination circuit (PSD) is used to distinguish the two
kinds of pulses with different fluorescent decay times.
In front of the primary detector a plastic scintillator with thickness 2 mm 
is used to distinguish charged particles. 

  In high sensitivity observation, the angular resolution is very important to 
avoid the confusion of point sources. The principal objective of the HXMT 
is to obtain the map of hard X-ray sky by the aid of the direct demodulation 
technique. The designed 10$'$ angular resolution and 1$'$ source location 
precision of the HXMT can help to accomplish the identification of sources in 
radio, optical and soft X-ray bands, on the other hand, images of the extended 
sources like the Galactic plane, the Galactic halo, clusters and SNRs can also 
be obtained. Imaging for a wide field and increasing exposure time during 
scanning observation require a wide field of view (FOV). However, a too wide 
FOV will damage the resolution of imaging, precision of source location and 
sensitivity of the instrument. After a comprehensive consideration of the 
requirement of imaging and flexibility of technology, the view field of the 
HXMT is chosen to be 5$^{\circ}\times5^{\circ}$ (FWHM), which consists of 18 
collimators with non-symmetric FOV of 5$^{\circ}\times0^{\circ}.5$ (FWHM) 
placed with a cross angle of 10$^{\circ}$ one another.
\begin{figure}
\psfig{figure=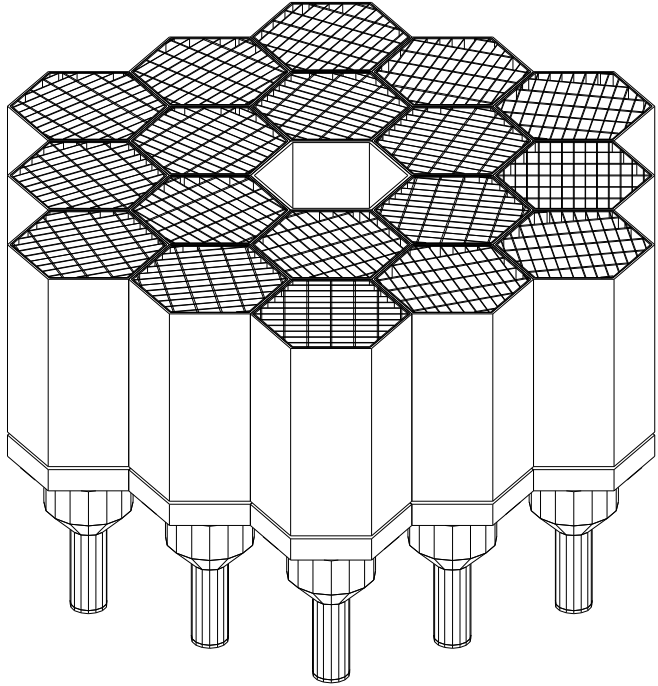,width=10cm,angle=0,%
bbllx=50pt,bblly=220pt,bburx=520pt,bbury=650pt,clip=}
\vspace{5mm}
\caption{Schematic diagram of the main HXMT detector}
\label{picture}
\end{figure}

  The spatial resolution ability of the HXMT was studied with the high precision 
reconstruction for part celestial region within the FOV based on direct 
demodulation method. The data of deep scanning survey in a
 $6^{\circ}\times6^{\circ}$ region which contains a strong source generated 
by computer simulation. The scan mode is: the angular differences between 
two sequential pointing directions is $0^{\circ}.5$ within a 
$1^{\circ}.5\times1^{\circ}.5$ region near the source and 1$^{\circ}$ 
in other region. The total observation lasts 2 hours. The background intensity 
is assumed to be 0.02 cm$^{-2}$s$^{-1}$ and the source has an intensity of 
$1.7\times10^{-2}$ cm$^{-2}$s$^{-1}$. From the resultant image of the point 
source by direct demodulation process the angular resolution (FWHM) of the HXMT 
can be estimated as $<2'$ and the accuracy of source location is much better 
than $1'$.

  The effect of the stability of satellite attitude on the HXMT 
spatial resolution was further studied. In a simulation the pointing axis 
moves along a random direction with a speed of $0^{\circ}.005$/s for each 
pointing observation. The mean pointing direction of each crystal with respect 
to the mean pointing direction and the number of photons detected 
in each crystal during the observation are all calculated. The difference 
between the real direction of the source and its calculated value from 
the direct demodulation map is only about $0'.4$, and the angular resolution of 
the point source is about $5'$ (FWHM). 

 The background was estimated with Monte-Carlo simulations and according to 
experiential results (Dean et al. 1989; 1990).
The solid angle of $5^{\circ}\times0^{\circ}.5$ 
collimator of HXMT is 7 times smaller that of IMAGER/INTEGRAL, 
which is much beneficial to reducing background of hard X-ray photons.
The estimated total background rate in the energy
range $10 - 200$ keV is 0.02 cm$^{-2}$s$^{-1}$. 
The following formulas are used to calculate the minimum detectable flux
of a pointing observation: 
\[ F_{min} \mbox{(cm$^{-2}$s$^{-1}$keV$^{-1}$)} = (2 n_{\sigma}/\varepsilon)
\sqrt{B/AT\bigtriangleup E} \] 
for continuum spectrum and 
\[ F_{min}\mbox{(cm$^{-2}$s$^{-1}$)} = (2.4n_{\sigma}/\varepsilon_{p})
\sqrt{B\delta E/AT}\] for narrow lines.
Where $n_{\sigma}$ is the significance, $B$(cm$^{-2}$s$^{-1}$keV$^{-1}$) 
background level, $A$(cm$^{2}$)  detector area, $T$(s)  observation 
duration, $\varepsilon$  efficiency, $\bigtriangleup E$(keV) widths 
of energy bands, $\varepsilon_{p}$ the detection efficiency of full energy 
peak and 
$\delta E$(keV)  the FWHM of the narrow line.
It has been estimated that the expected sensitivity of HXMT at 100 keV 
is three times better than that of IMAGER/INTEGRAL.

HXMT can make observation in three different modes: full-sky survey, deep scanning
observation of a selected region, and pointing observation.   
In the sky survey mode, the satellite attitude is three-axis stabilized  
with respect to the earth.  
The survey is carried through the motion of the satellite in its 
orbit and precession of the orbital plane. For a satellite with a low
circular orbit at the altitude of 550 km and an inclination of 43$^{\circ}$, 
the scan survey is divided into three phases with roll angle being $0^{\circ}, 30^{\circ}$ 
and $-30^{\circ}$, separately.
The above scan observations will take six months to cover most of the sky.
The three-phase scanning observation mode ensures that some regions can be observed up to 
six times. 
After the sky survey, HXMT will start to pointing 
observation of some objects and deep scanning observation of selected sky regions.  
In this phase the attitude
of the satellite is controlled in the 3-axis stabilized mode with respect to stars in the sky.
With the 3-axis stabilized mode guided by the star image,
HXMT will 
perform scanning monitoring of the sources in the galactic 
plane several times for about every ten days. The time to finish a scan of the 
galactic plane is the orbital period of the satellite, in order not to be 
affected by the earth occultation.

  The key performance parameters of HXMT mission are listed in Table 1.
%\newpage
\begin{center}
Table 2: Key Performance Parameters of HXMT Mission  
\end{center} 
\begin{tabular}{ l l} \hline  
 Energy Range	   &  10--200 keV \\ %\hline
 Energy Resolution &  $\sim 18\%$ @ 60 keV 	\\ %\hline 
Angular Resolution &  $\leq 10'$   		\\ %\hline
 Source Location ($20\sigma$ source) &  $\leq 1'$ 	\\ %\hline
Sensitivity ($3\sigma$, in $10^{5}$s @ 100keV) 
 & $3\times 10^{-7}$ cm$^{-2}$s$^{-1}$keV$^{-1}$ (continuum)\\ 
      & $1\times10^{-5}$ cm$^{-2}$s$^{-1}$ (narrow line)\\ %\hline
Orbit &  Altitude: $\sim 550$ km circular\\
      &  Inclination: $\sim 43^{\circ}$\\ %\hline
Attitude & Three-axis stabilization\\
         & Control precision: $\pm 0^{\circ}.25$\\
         & Measurement accuracy: $\pm 0^{\circ}.01$ \\ %\hline
Data Rate & $\sim 30$ kbps \\ %\hline
Mass & Science instrument: $\sim 600$ kg \\
     & Total payload: $\sim 1400$ kg \\ %\hline
Nominal Mission Lifetime & 2 years \\ \hline
\vspace{5mm}
\end{tabular}

The HXMT mission was proposed in 1994 and selected as one of  the Major State 
Basic Research Projects in China in April, 
2000 and funded by the
Ministry of Science and Technology of China, Chinese Academy of Sciences and 
Tsinghua University. 
The implementation of the HXMT project is a collaboration between the 
Chinese Academy of Sciences (CAS)
and the Tsinghua University.     
With the fund the detector system and the
prototype of the payload are under construction.
It has also been listed as a candidate of the state scientific 
satellites of China. International collaborations on this project are welcome.
In 10 - 200 keV hard X-ray range, HXMT mission with its unprecedented 
sensitivity and imaging ability is expected to play an important role 
in space astronomy.

\section{Discussion: Direct Approach to Inversion} 
Sufficiently using the information on the observed object is of
great importance in dealing with the reconstruction problem. 
The observation equation system
\[\hspace{30mm} \sum\limits_{i=1}^{N}p(k,i)f(i)=d(k)
\hspace{10mm} (k=1,...,M) \hspace{31mm}(1) \]
or   
\[\hspace{60mm} Pf=d \hspace{56mm}(2) \]
contains all information about the object $f$ in the observed data $d$ 
obtained by the observation with an instrument $P$. The maximum-entropy
method (ME) is a widely used method of reconstructing object $f$ from 
observed data $d$. ME is to choose a solution $f(i) (i=1,...,N)$ from
all satisfying the statistical criterion

\begin{equation} \sum_{k=1}^M\{[\sum_ip(k,i)f(i)-d(k)]/\sigma_k\}^2=M 
\end{equation}     
by the condition that the information entropy
\begin{equation} -\sum_if(i)\log f(i)=\max \end{equation}
Many informations on the object $f$ are lost when using the single 
equation (9) instead of the equation system (1) containing $M$ equations, 
that infinite intensity distribution $f$ can satisfy Eg. (9).
The maximum-entropy condition (10) helps us to pick out a most smooth one
from them, but not necessary the best one. Another widely used method is
the cross-correlation technique, which takes the cross-correlation 
distribution $c$ of the data and the modulation pattern to estimate the
object intensity
\begin{equation}
f\propto c=P^Td \end{equation}
From the correlation equations 
\[ P'f=c   (5) \]
one can see $c\ne f$, the cross-correlation $c$ is just an image of the
 object $f$ through a modulation (distortion) of a certain imaging instrument
with a point spread function $P'$. The cross-correlation reconstruction
completely ignores the information included
 in the correlation equations, still does not make full use of
the information on the object obtained by the observation. 
In general the two kinds of inversion methods have no significant 
difference in their results, the resultant resolution are limited
by the intrinsic resolution $\Delta_0$ of the instrument
\begin{equation}
\Delta_{CC}\approx \Delta_{ME}\approx \Delta_0 \end{equation}   
      
The direct decording approach, reconstructing 
objects by iteratively solving observational equations or correlation 
equations under nonlinear constraints, each iterative calculation 
is directly based on the modulation equation. Obviously the direct approach
can use the information 
containing in the equations more sufficiently  
than any indirect one through a statistical criteria or cross-correlation
transformation and then can obtain much better resolution 
$\Delta_{DD}\ll \Delta_0$. 

With conventional techniques the resolution is limited by the
intrinsic one, it does not matter how sensitive the observation
is.  For improving the precision and resolution 
of an experiment, people are usually devoted 
to improving the intrinsic resolution by constructing
high-precision instrument, more and more complicated and high-priced.
It is a most difficult task to conduct an experiment with both 
high resolution and high sensitivity.
Under the direct inversion technique, the resultant
resolution $\Delta_{DD}$ is not only dependent on the instrumental 
intrinsic resolution $\Delta_0$,
but also dependent on the amount $S$ of observed signal, the ratio of
signal to noise $SNR$ and modulation pattern $M$
\begin{equation}
\Delta_{DD}=f(\Delta_0, S, SNR, M) \end{equation}
Applying the direct demodulation technique to space instrument design, 
we can realize high resolution imaging with a low resolution 
or even non-position-sensitive detector, and can conduct observations
with both high resolution and high sensitivity by improving the resolution
ability through increasing the detection sensitivity, as shown 
by the expected performance of HXMT mission.  

The main difficult of the direct methods is the illness of solution causing
by errors in data and other uncertainties in modulation equations. Introducing
physical constraints to make nonlinear control in
the iterative process of solving modulation equations is a key point
to make the direct demodulation realizable, and, furthermore, stable,
convergent and global optimal. Benefited from the physical constraints
in iterates the effect of errors can be depressed effectively
in the demodulation process. As a consequence, the direct demodulation can
get images with much less false appearances and be capable of 
reconstructing not only discrete sources, but also extended features as well.

\end{document}